\magnification = \magstep1
\pageno=0
\hsize=15.truecm
\hoffset=0.5truecm
\vsize=22.0truecm

\output={\plainoutput}
\pretolerance=3000
\tolerance=5000
\hyphenpenalty=10000    
\newdimen\digitwidth
\setbox0=\hbox{\rm0}
\digitwidth=\wd0

\def\footnoterule{\kern-3pt \hrule width \hsize \kern 2.6pt
\vskip 3pt}
\def\cl{\centerline}
\def\ni{\noindent}

\def\msun{M$_{\odot}$}

\def\vs{\vskip 11pt}

\def\solar{\ifmmode _{\mathord\odot}\else $_{\mathord\odot}$\fi}

\font\ksub=cmsy7
\def\teff{T$_{\kern-0.8pt{\ksub e\kern-1.5pt f\kern-2.8pt f}}$}

%
\vs\vs\vs
\vs\vs\vs
\vs
\vs
\vs
\cl{\bf RADIATIVELY-DRIVEN COSMOLOGY}
\vs
\cl{\bf IN THE CELLULAR AUTOMATON UNIVERSE}
\vs\vs
\cl{Robert L. Kurucz}
\cl{Harvard-Smithsonian Center for Astrophysics}
\vs
\cl{May 13, 2006}
\vs
\eject
\vs\vs\vs
\cl{\bf RADIATIVELY-DRIVEN COSMOLOGY}
\vs
\cl{\bf IN THE CELLULAR AUTOMATON UNIVERSE}
\vs\vs
\cl{Robert L. Kurucz}
\vs
\cl{Harvard-Smithsonian Center for Astrophysics, 60 Garden St, Cambridge, MA 02138}
\vs\vs\vs
 
\cl{ABSTRACT}\vs
 
This is an updated version of my paper ``An outline of radiatively-driven
cosmology" (Kurucz 2000).  Here the Big Bang universe is replaced by a
finite cellular automaton universe with no expansion (Kurucz 2006).  The 
Big Bang is replaced by many little bangs spread throughout the universe
that interact to produce the initial perturbations that form Population III 
stars, globular clusters, and galaxies, but no large-scale structure.
These perturbations evolve into the universe as we now observe it.  
Evolution during the first billion years is controlled by radiation.  
Globular clusters are formed by radiatively-driven implosions, galaxies 
are formed by radiatively-triggered gravitational collapse of systems of 
globular clusters, and voids and the microwave background are formed by 
radiatively-driven expansion.  After this period most of the strong 
radiation sources are exhausted and the universe relaxes into gravitational 
old age as we know it.  To relieve the boredom we present the results of 
gedanken experiments (Kurucz 1992) in a traditional, linear, chronological 
sequence in the hope of stimulating research on the many topics considered.

\vs 
\ni Subject headings: cosmology --- stars: Population III --- stars: Population II ---
clusters: globular --- galaxies: evolution
\vfill
\eject
\cl{THE CELLULAR AUTOMATON UNIVERSE}
\vs
I repeat here the sections from my paper on ``Elementary physics in the cellular
automaton universe" (Kurucz 2006) about the intitial conditions that 
evolved into the present day universe. 

\centerline {Black Holes}
Black holes are easy to explain without singularities.  Adding mass to a
neutron star causes the neutrons to collapse into boson di-neutrons with 
spins anti-parallel,  n + n = udd + udd $\to$ uddudd = di-n.  The neutron 
star becomes a di-neutron star.  This can be a gradual transformation, not 
a catastrophe.  If matter is added slowly, the neutron star becomes an 
invisible black hole and continues to grow until the fermion nature of the 
quarks limits the compression.

Continuing to add mass to a di-neutron star causes the quarks within the 
di-neutrons to pair with spins opposed so they become di-quark bosons with 
0 spin, di-n = uddudd $\to$ uu-dd-dd = di-q-di-n.  The di-neutron star 
becomes a di-quark-di-neutron star which is a super-massive black hole. 

     A cellular automaton has to have a density limit to prevent overloading
the ``computation" in a small volume.  The simplest cutoff is to make 
gravity repulsive at high density to automatically blow apart dense
concentrations.  This could be built into the cellular rules for each particle.
\vs
\centerline {Antimatter}
There is a missing anti-matter problem if this universe began in a Big Bang
of radiation.  Starting with radiation implies that all primordial particles 
were made by pair production as the universe cooled.  If pair production
does not dominate, then matter and antimatter do not have to balance.

The idea of a di-quark-di-neutron black hole suggests that our universe 
``started" from a collection of ultra-massive black holes 
statistically uniformly distributed throughout the cellular automaton.
For example, the initial state might be $10^{12}$ to $10^{13}$ 
$10^{13}$-to-$10^{14}$-solar-mass black holes with average separation 
less than a megaparsec.

At the first tick the density in the ultra-massive black holes exceded
the density cutoff so gravity was repulsive.  The ultra-massive black
holes expanded at sub-light speed.  The di-quark-di-neutrons expanded and 
became di-neutrons.  The di-neutrons
expanded and became neutrons.  The neutrons expanded and became protons and
electrons and anti-electron-neutrinos, and deuterons, and alpha particles, 
etc.  The initial number of neutrons was fixed.  The proton number, the
electron number, and the anti-electron-neutrino number are equal.  Subsequent
pair production does not affect the baryon (neutron+proton) total.  There is
no antimatter problem.  The universe is fundamentally biased toward matter.

The statistical equilibrium and the formation of nuclei led to different 
abundances, including heavier nuclei, and different properties than we are 
used to from a Big Bang prediction.

\vfill
\eject
\cl{FORMATION OF PERTURBATIONS}
\vs
Figure 1 shows the randomly positioned cold ultra-massive black holes that 
fill the universe with mean spacing less than one megaparsec.  Each black 
hole has twelve or thirteen neighbors with which it will collide once it 
expands.  At the first tick of the cellular automaton the particles take 
their first step and the black holes expand in ultra-massive 
supernova-like explosions.  The massy particles move outward at less 
than the speed of light.  They interact, heat, and generate photons and 
neutrinos that move outward at the speed of light into the emptiness of 
the cellular automaton.  After a million years they pass photons and 
neutrinos coming in the opposite direction from the neighbors.  The massy 
electrons, protons, neutrons, He, Li, Be, B, C, N, O, et cetera travel 
much more slowly.  It takes millions of years before the massive shells 
collide with their neighbors.

The expanding shells lose material in the backward direction.  They radiate
thermally in all directions and cool.  The radiation eventually, after
one or two million years, reheats all the neighbors.  Each expanding shell
is hit by twelve or thirteen radiative precursors in sequence from its
neighbors.  All the background radiation is absorbed.  There is no 
background radiation left from the initial universe.

Continuing with Figure 1, the expanding shells collide, pancake, and facet.
Eventually amoeboid condensations fill the interstices with the material from 
twelve or thirteen successive collisions from different directions.  The 
amoebas are filled with perturbations varying in size from few-hundred-solar-mass 
Population III-star-size, to one- to ten-million-solar-mass globular-cluster-size 
perturbations.  Outside the condensations are the low density regions from 
whence the shells expanded.  The perturbations on the amoeboid surfaces radiate 
outward into the low density regions and cool rapidly.  Perturbations in the 
interior of the amoebas must radiate into their neighbors and absorb radiation 
from their neighbors, so they cool more slowly. 

\vs
\vs
\cl{FORMATION OF POPULATION III STARS}
\vs

Small, few-hundred-solar-mass perturbations cool faster than larger perturba-
tions because they have a large surface to volume ratio.  Perturbations 
inside the condensed regions are illuminated on all sides by radiation from 
other perturbations.  Perturbations on the surface of the amoeba radiate 
efficiently into an empty hemisphere.  Hundreds of thousands of diatomic 
hydride and hydride ion lines can transfer energy from the ultraviolet 
electronic bands to the infrared vibration-rotation bands, even at low 
abundances.  Cooling and Population III star formation become easy. 

\vfill
\eject
\cl{FORMATION OF GLOBULAR CLUSTERS}
\vs
 
     Massive Population III stars are superluminous.  They radiate
about 10$^{53}$ ergs (10$^{51}$M$_{*}$/\msun) in 10$^{6}$ years and then 
explode as supernovas.  That is enough radiation for serious construction
projects using radiatively-driven implosions.  The words ``radiatively-driven 
implosion" and even the concept used to be classified.  When they
were declassified, astrophysicists who work at the weapons laboratories  
could apply them to bigger problems.  Simple examples of radiatively imploding 
a bump on the surface of a Population I molecular cloud and of radiatively 
imploding a small Population I cloud between two hot stars have been presented 
in a series of papers by Sandford, Whitaker, and Klein
(Sandford, Whitaker, and Klein 1982; 1984; Klein, Sandford, and Whitaker 1983).

     Figure 2 shows that a cloud between two O stars can
be compressed into a dense globule in 30000 years.  Figure 2 also shows open
cluster formation by successive generations of implosions.  Sandford, Whitaker,
and Klein never extrapolated that idea to the formation of a globular cluster,
but I do.  A bump on a cloud that is illuminated by an O
star is imploded into a dense globule by compression both from the front,
and also by compression from the sides by scattered light or by dust pushed 
by scattered light.  

     These examples can be generalized to gas clouds of any population
illuminated by a hot star or stars.  If there is an opacity that increases non-linearly 
with density, the bumps on the cloud surface will be compressed and become
optically thick while the surrounding gas is still optically thin and is
still scattering the starlight in all directions, including behind the
perturbation, as in Figure 3.  The perturbation is cut off from the cloud
face as a collapsing, rotating globule.  The globule is further compressed 
by the radiation from the star and radiates excess energy from the unilluminated
side that pushes the cloud away.  The mass must be large enough so that the 
unilluminated side is
in quasi-hydrostatic equilibrium or collapsing.  Otherwise the globule
loses mass in a wind until it evaporates.  The collapse continues into a 
nuclear burning star.  The mass of the globule and the mass of the star are
determined by the optical depth scale of the perturbation in the cloud wall.  
The lower the opacity the greater the minimum mass required.  Any leftover 
material in the outer shell is driven inward.  The layering process repeats 
inward (an onion skin model) until all the matter in a large perturbation 
is formed into stars.
Four massive stars arranged tetrahedrally will radiatively implode a one- 
or even a ten-million-solar mass cloud into a globular cluster.  Additional 
hot stars accelerate the collapse.  Each massive star can illuminate more 
than one cloud at a time.  Globular clusters can be formed in any population, 
at any time.    

     Once the first Population III stars form, there is a hiatus of 10$^{5}$
to 10$^{6}$ years until they start to supernova.  Before there are supernovas
all gas is Population III gas.  The intitial Population III stars form 
additional Population III stars in and around their H II regions.  These new
stars are smaller than the original Population III stars because of
the radiatively-driven implosion.  The original stars did not have external 
help.  Because the lifetimes of Population III stars are so short, there is 
not enough time for larger perturbations to evolve before the Population III 
stars supernova.  All other matter in the universe is contaminated and 
becomes Population II material.

     Once the Population III stars start to supernova the gas becomes
more and more contaminated with metals and its opacity greatly increases.
Then radiatively-driven implosions form globular clusters of Population II 
stars that have a 
smaller minimum mass.  The number of supernovas increases, the minimum mass 
continues to go down until Population II stars form that are too small to 
supernova, 7 or 8 \msun.  These stars evolve over billions of years,
lose mass, and become white dwarfs.  
As the supernovas continue the abundances increase above 1/100 solar and
the minimum mass drops another factor of 10 or so to the point where these 
stars have not yet evolved off the main sequence.  Many supernovas early on
result in many low mass stars that evolve slowly, while few supernovas early 
on produce many high mass stars that evolve rapidly.

The stellar abundances and masses are determined by the number and proximity of
the supernovas.  The distribution function of these Population II masses is the
initial mass function.  The masses can range over the whole spectrum but
because the Population II material has higher opacity than the Population III
material, and because its collapse is helped along by external forces, the
masses are smaller than the Population III masses and can even be quite small.
However, the smallest Population II stars are still larger than the smallest
(future) Population I stars which form easily because of high opacity gas and
dust.  There are Population II M dwarfs but there are no Population II 
brown dwarfs.

     We can observe only supernovas that supernova, not the duds.  There can be
mass and angular momentum ranges where supernovas fail and produce oxygen
and other alpha-process elements but no iron and no heavy elements.  This is
especially likely for Population III stars because of their large masses and 
rapid evolution.  Duds could produce solar masses of alpha elements that would
contaminate the globular-cluster-size clouds and greatly increase the opacity 
without increasing the iron abundance.  Diatomic hydrides and oxides and their ions
have millions of lines and are very efficient at radiating away excess energy.   
With enough oxygen it is even possible to make water with its tens of millions
of lines.  When most supernovas are successful, the alpha enhancement is a factor 
of 2 to 3 and the Fe varies from 1/100 to 1/1000 solar.   These opacities are
high enough to produce K dwarfs.  When the condensations have relatively few 
successful supernovas that produce iron and many duds, the opacity becomes high 
enough to produce M dwarfs with alpha enhancements greater than 100 and Fe 
abundances 1/1000 to 1/10000 solar.  With many successful supernovas and many 
duds the opacity becomes high enough to form M dwarfs with Fe abundances  1/300 
solar but with alpha enhancements greater than 10.  M dwarfs with high alpha 
abundances have not evolved off the main sequences in the age of the universe 
and are invisible for all practical purposes.  

     The amoeboid pseudopods of the condensations have maximum exposure
and radiate most easily of all the condensed matter.  I suggest that the
pseudopods have so many dud Population III supernovas that most of the mass
is converted into globular clusters full of M dwarfs.  More than half of the
mass of the universe is sequestered in invisible M dwarfs.
 
     Globular cluster formation goes much more slowly within the amoeboid
main body of the condensation.  The stellar mass can vary from massive, greater
than seven solar masses, down to K dwarfs greater than one half a solar mass.

\vs
\vs
\cl{FORMATION OF GALAXIES}
\vs
 
     Asymmetries in the distribution of the Population III stars around each
globular-cluster-size perturbation produce a small, net globular cluster velocity.  
Since there are excess Population III stars at the surface of the condensed
region, the globular clusters near the surface of the central condensation 
will be accelerated inward and will have velocities inward on the order of a 
fraction of a km s$^{-1}$.  This is the radiative trigger that leads to the 
gravitational implosion (violent relaxation) of the systems of globular 
clusters into elliptical galaxies.  Figure 4 shows a schematic calculation 
of such violent relaxation.  As part of the collapse, the most energetic 
globular clusters are ejected into the low density regions.  As galaxy-size 
perturbations have no symmetry, they have angular momentum and they spin up 
as they collapse.  The globular clusters in the pseudopods either form their 
own galaxies around the central small cluster of galaxies, or they form a 
halo around it, or both.
 
     At this point the universe is filled with small clusters of elliptical 
galaxies surrounded by halos of M-dwarf elliptical galaxies and globular 
clusters, all surrounded by low density regions.  The low density regions are 
less than a megaparsec in diameter.  All of the globular clusters in these 
elliptical galaxies have approximately the same age.  
The globular clusters collide and gain internal
energy and rapidly disintegrate.  By today 99.9\% of them have disintegrated.
The clusters that are left are not typical or representative of the properties
of the initial ensemble.  They are the cold tail.  They are not pure, having 
added and lost stars through their whole lives.  The current members of one of 
these globular clusters are not necessarily siblings, coeval, or even 
Population II.  There can be dark globular clusters in which all or almost 
all the stars are neutron stars and white dwarfs.  There are globular 
clusters of M dwarfs that have not yet evolved.

     As they evolve, the elliptical galaxies with mainly high mass stars 
become opaque from the many supernova remnants that fill their 
halos.  These opaque galaxies make the universe opaque.  Elliptical galaxies 
with lower mass stars that do not supernova are transparent and faint 
until their stars evolve up the giant branch and lose mass.
\vfill
\eject
     The low density regions contain globular clusters and individual 
stars that originally ranged from massive to M-dwarf, neutron stars, 
black holes, and Population II gas blown out of galaxies and clusters and 
from mass loss.  The low density regions also collect the alpha-enhanced 
M dwarfs that escape from the galaxy cluster halos. The low density regions
are transparent but they can absorb radiation in lines.

     Figures 5 through 12 schematically describe galactic evolution.

     If the initial mass functions of the globular clusters that form
an elliptical galaxy have almost all low mass stars, the galaxy remains
an elliptical galaxy forever.  These galaxies have low luminosity until
the giant branch is strongly populated.  A few, more massive, stars 
lose enough mass to fill the galaxy with the tenuous gas that produces 
the Lyman $\alpha$ forest.
 
     If the initial mass functions of the globular clusters have mostly
high mass stars, the elliptical galaxy evolves into a spiral galaxy.
Supernova remnants and the mass lost by intermediate mass supergiants
collapse into a bulge and a disk, which spin up.
 
     An intermediate case produces an irregular or ``young" galaxy.  

     When there is a significant high mass tail, after some 20 million years,
the whole elliptical galaxy fills with supernovas and supernova remnants.  
The galaxy fills with jumbled magnetic structures.  The galaxy 
becomes opaque.  The supernova remnants and the magnetic structures cannot 
orbit because of their large collision cross-sections.  They collapse into 
a central bulge with a quasar at the center. 

Since the 
supernova remnants have high abundances, the bulge gas has high abundances 
and must form high abundance stars.  This can happen both in galaxies that 
are today elliptical or spiral.  These initial quasars continue to be powered 
by infall of gas that is blown off intermediate mass stars when the stars 
climb the giant branch.  This gas is low abundance Population II gas.  It 
dilutes the supernova remnant gas.  This gas forms the disk of spiral galaxies 
so that the first stars in the disk have abundances initially lower than bulge 
abundances.  
 
     The activity that we have been describing takes place in the first
10$^{9}$ years.  The time scales are set by orbital and collapse times, and
by stellar evolutionary time scales.  It takes, say, one orbital time
to form the bulge and quasar, and a few orbital times for the mass loss
infall to form the disk.
 
    Since the disk is formed from mass-loss material from Population II stars
in the halo, the mass of the disk gives a lower limit to the mass of the
one- to six-solar-mass primordial Population II stars in the halo and to the
number of white dwarfs.  Each star loses its own mass less the mass of a
white dwarf.
 
    Since the quasar and bulge are formed from 
supernova remnants, the mass of the central object and bulge (less the
equivalent volume of halo stars) give a lower limit to the mass of the, say,
7 solar mass and greater primordial Population II stars in the halo and
to the number of neutron stars and black holes  Each star loses its own 
mass less the mass of the neutron star or black hole.

\vfill
\eject
\cl {FORMATION OF BULGES, QUASARS, DISKS, AND BARS}
\vs

     The massive initial Population II stars blow out most of their mass
in a supernova explosion and leave a neutron star or black hole behind. 
The neutron stars, black holes, gas, and magnetic field remnants continue 
in their halo orbits.  The orbits cross the plane of the galaxy and many 
of the orbits pass near the center.  Since the gas and magnetic fields have 
large collision cross-sections, they interact near the center.  The 
collisions cancel part of each component of angular momentum to produce a 
bulge that is a small model of the original elliptical galaxy.  Since the 
bulge material is highly enriched in metals from the supernovas, and
since it is continually shocked and compressed by collisions, it has high
opacity and readily produces new massive second-generation Population II stars that
rapidly implode the remaining bulge material into globular clusters with high
metal abundances.  In general, all of the second generation Population II stars
are less massive than the corresponding initial Population II stars because
the higher abundances allow the gas to radiate and collapse more readily.

     Black holes, including the Population III black holes, and neutron stars
orbit through all this condensing material in the bulge and they accrete some 
of it.  They are much more likely to have gravitational interactions among
themselves and with other halo stars as they pass through the bulge than when 
they are out in the halo.  They tend to relax toward the center and to
agglomerate with halo stars, bulge stars, and each other.  Some of the black 
holes become big enough to dominate their environment. The largest black hole
eventually ends up at the center of the galaxy and is orbited by the other 
black holes.  The black holes form a second smaller model of the original 
elliptical galaxy and of the bulge because they have cancelled part of each 
component of their angular momentum, Figure 11.  The black holes that orbit the 
central black hole are ``Kerberean" black holes.  They guard the central black hole 
by sweeping up incoming material.  They collect the mass and angular momentum
of all the stars, gas, low-mass black holes that fall toward the center of the 
galaxy.  When material falls toward the center faster than the Kerberean black 
holes can sweep it up, 
it falls into the central black hole and produces a flaring that is called a 
quasar.  If only a small amount of material falls into the center, it produces 
a flaring called an active galactic nucleus.  When there is a collision between
galaxies, material from the other galaxy can fall toward the center and also
produce flaring.  

     The quasar can be formed only during the first few orbital periods of a 
galaxy because the massive stars that evolve into black holes have short 
lifetimes and because the remnant material is collected in the bulge when it
tries to orbit through.

     The mass of the bulge is proportional to the number of initial Population
II stars that supernova.  There is a minimum startup value because there have
to be enough remnants colliding at the center to start the agglomeration.
The sum of the mass of the central black hole and the Kerberean black holes 
is also roughly proportional to the number of Population III supernovas and 
intitial Population II supernovas.  Massive black holes ingest a significant 
fraction of the metals that were produced by the Population II 
supernovas.  Only a fraction of the black holes end up in the
bulge; the remainder are still in halo orbits.  

     The halo is depleted of material that orbits through the bulge.  Material
in less eccentric orbits away from the bulge is most likely to collide and collapse 
as it crosses the equator.  After one hundred million years, mass loss from stars 
that do not supernova, with masses less than seven or eight solar masses, begins.
The low abundance mass-loss material, combined with remaining high abundance 
supernova remnant material, loses some angular momentum in all components in the 
collisions, but most in the x and y directions, so that it spins up. It collapses 
into an oblate spheroid, a thick disk, with an empty center that contains the bulge.  
Its diameter is much less than that of the halo.  As lower mass stars evolve, 
the mass loss continues, and the disk grows and become more efficient at 
collecting mass.  The newer materal forms a thin disk in the equitorial plane 
of the elliptical galaxy.  As material continues to fall in (from evolved K 
dwarfs at present) the disk grows in size and structure.

    The bulge repeats the evolution of the halo.  The more massive stars
supernova and the material falls toward the center.  Orbits that pass
near the center are depleted.  As the lower mass stars evolve, the lost mass
forms an oblate spheroid at the equator of the bulge with an empty center.
If the disk has formed strong spiral arms, the spheroid becomes triaxial and
forms a bar.  The spheroid merges with the disk.  The Kerberean black holes 
continue to orbit the central black hole and continue to grow.

\vs
\vs
\cl{FORMATION OF VOIDS AND THE MICROWAVE BACKGROUND}
\vs
 
      Voids and the microwave background are remnants of the radiatively-driven 
expansion produced by quasars.  Primordial galaxies manufacture
a tremendous amount of radiation.  Any galaxy that is a spiral now originally
had most of its mass in massive stars.  A 10$^{12}$ \msun\  spiral galaxy
produces, say, 10$^{11}$ supernovas yielding 10$^{62}$ ergs.  The precurser
stars radiate even more during their lifetimes, say 10$^{63}$ ergs.  There
might be 3x10$^{11}$ intermediate mass stars that radiate 10$^{63}$ ergs and
end up as white dwarfs.  In addition the quasar itself produces 10$^{46}$ to
10$^{47}$ ergs s$^{-1}$ for say 3x10$^{8}$ years or about 10$^{63}$ ergs.  There is
also a great deal of energy from the collapse that heats the gas and is eventually
radiated away, partly by the quasar.  Integrating over the first billion years, 
letting one-half the large galaxies be spirals, it is easy to produce 10$^{51}$ 
ergs \msun$^{-1}$\ averaged over all galaxies.
Neutrinos produced by the supernovas add up to a similar amount of energy.
 
      During much of the first billion years the elliptical galaxies that 
become spirals are filled with debris from supernovas and mass loss.  They
are optically thick.  They self-absorb most of their own radiation and they
absorb all infalling radiation from other galaxies.  After the bulges and 
quasars collect most of the orbiting material, much of the radiation
generated by the galaxies can escape, but the galaxies are still optically 
thick to radiation passing all the way through the galaxy from the outside,
Figure 14.
The galaxies still absorb all the incoming radiation.  The universe is still 
optically thick.  In this situation radiation pressure 
pushes the galaxies apart.  Random dense groups of large galaxies with many 
bright quasars push themselves apart and form voids.  They and smaller 
galaxies expand outward in a shell.
The transparent galaxies are dragged along by gravity.  When the quasars
in smaller galaxies come on line those galaxies are already moving outward 
in a shell.  Their radiation further increases the size of the shell and the
void.  The voids collide, force thinning and opening of the shells, and merge.  
The collisions produce streaming and galaxy clustering in the walls of the
voids.  The quasars eventually burn out.  The gas in the halos of spiral 
galaxies collapses inward and begins to form the disk, leaving the halo 
transparent and filled with dead and unevolved stars.  The elliptical galaxies 
are transparent and have halos filled with unevolved stars.  The universe 
becomes optically thin.  It becomes possible to see for great distances 
through halos and outer halos of the intervening galaxies.  We can see a few 
of the fading quasars as they complete their job of restructuring the universe 
through radiatively-driven expansion.

     Reg\H{o}s and Geller (1991) have shown that some of the small, low-density
expanding regions in a uniform background will continue to expand gravitationally
in an expanding universe, Figure 15.  They form voids that collide and merge.  
The collisions produce large galaxy clusters, streaming in the void walls, and
eventually large scale structure and motions that we see today.  The figures 
will be qualitatively the same if the universe is not expanding, but instead 
radiation in doing the work.

     The hot radiation in the ultraviolet and visible that was re-radiated in 
the infrared was absorbed by galaxies and pushed them apart.  That radiation
was eventually re-radiated in the infrared beyond 100 $\mu$m.  The radiation 
given off by more than 
10$^{11}$ quasars became the background that we now see redshifted into the 
microwave.  It is smooth because it averages over so many galaxies and
over a long time interval.  The apparent temperature turnover is caused by 
absorption from gas and dust at the quasar redshift that increases blueward.
The background is modulated by the pattern of the voids breaking open and 
becoming optically thin.  The pattern has nothing to do with primordial 
cosmology.  Part of the motion detected as the microwave background
dipole is a remnant of radiatively-driven streaming from void formation.

\vfill
\eject
\cl{FIGURES}
\vs
Figure 1.  The universe is a finite cellular automaton initially 
``started" from a collection of ultra-massive black holes 
statistically uniformly distributed throughout the cellular automaton.
For example, the initial state might be $10^{12}$ to $10^{13}$ 
$10^{13}$-to-$10^{14}$-solar-mass black holes with average separation 
less than a megaparsec.  The black holes explode and collide with
twelve or thirteen neighbors to form amoeboid galaxy-cluster-size 
condensations filled with globular-cluster- and Population III-star-size 
perturbations.  Perturbations in the pseudopods eventually form M-dwarf 
halos around the central condensations.

Figure 2.  Simulations of radiatively-driven implosions of Population I
clouds indicate the plausiblity of forming a globular cluster by surrounding
a cloud with hot stars.  

Figure 3.  Radiation from a massive star radiatively implodes bumps on a
cloud surface into stars.
  
Figure 4 qualitatively demonstrates that small radiative accelerations
are sufficient to trigger the collapse of a universe full of globular
clusters into a universe full of elliptical galaxies.  I borrowed the program
from Reg\H{o}s that she used to model void formation (Reg\H{o}s and Geller 1991).
The universe is periodically tesselated into cubes with constant density of 
globular clusters, 128**3 per cube.  Each cube is subdivided into 8
parallellopipeds as shown in the upper left.  This is an arbitrary choice
intended not to look like galaxy precursers.  All the surfaces of all the 
parallellopipeds are given a small inward velocity as would be produced
by excess supernovas at the the surfaces.  The initial condition is zero  
gravitational force.  The small motion of the surface globular
clusters is enough to cause violent relaxation into a galaxy, except in one
case where neighboring galaxies cause the smallest object to disintegrate
and then assimilate its remains.  The most energetic globular clusters are 
sprayed outward and escape from the galaxy cluster.  They are not visible in 
the figure.

Figure 5.  Schematic evolution of galaxy of 1/2 \msun\  stars.

Figure 6.  Schematic evolution of galaxy of 1 \msun\  stars.

Figure 7.  Schematic evolution of galaxy of 10 \msun\  stars.

Figure 8.  Schematic evolution of galaxy with distribution function peaking 
at 2/3 \msun\  stars.

Figure 9.  Schematic evolution of galaxy with distribution function peaking 
at 1 \msun\  stars.

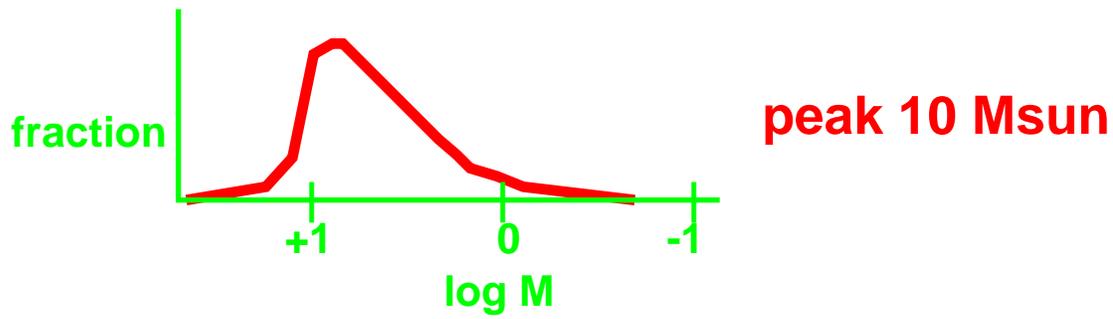
Figure 10.  Schematic evolution of galaxy with distribution function peaking 
at 10 \msun\  stars.

Figure 11.  Evolution of our galaxy.

Figure 12.  Isolated galaxy classification as a function of galaxy mass 
and of stellar mass distribution function peak.  All spiral and irregular 
galaxies that have not been damaged by collisions or interactions have 
large, massive, elliptical halos.

Figure 13. There is a tremendous radiation pressure from quasars because
all their radiation is absorbed by other galaxies, the galaxies with halos
filled with supernova remnants.  Random groups of quasars generate enough 
overpressure to form voids.  Transparent galaxies are dragged along by 
gravity.

Figure 14. Reg\H{o}s and Geller (1991) showed that starting with a uniform
density universe, one could evolve voids and large scale structure by
removing half the matter from small spheres and redistributing it in
expanding shells.  Even if the universe is not expanding the radiative
forces are so strong that they produce the same result.

Figure 15.  Table of contents of our galaxy.  In addition,
our galaxy is surrounded by an external halo of about 10$^{13}$ high-alpha
M dwarfs at distances greater than 100 kiloparsecs.  There are between one 
and ten M dwarfs per square arcsecond on the sky.  Their radiation is too 
faint to be detected.  Actually the whole local cluster is imbedded in an 
M-dwarf halo.  
\vs
\vs
This work was supported, in part, by many NASA grants from 1974 through 2005.
\vs
\vs
\cl{REFERENCES}
\vs
 
\ni Klein, R.I., Sandford, M.T.,II, \& Whitaker, R.W. 1983, ApJ, 271, L69
 
\ni Kurucz, R.L. 1992, Comments on Astrophysics, 16, 1-15.
 
\ni Kurucz, R.L. 1995, ApJ, 452, 102-108.
 
\ni Reg\H{o}s, E. \& Geller, M.J. 1991, ApJ, 377, 14-28.
 
\ni Sandford, M.T.,II, Whitaker, R.W., \& Klein, R.I. 1982, ApJ, 260, 183-201.

\ni Sandford, M.T.,II, Whitaker, R.W., \& Klein, R.I. 1984, ApJ, 282, 178-190.
 
\vfill
\eject
\end